\documentclass[11pt,twoside]{article}
\usepackage{asp2014}

\aspSuppressVolSlug
\resetcounters

\begin{document}

\title{Flares from coalescing black holes in the centimeter-wavelength transient sky}
\author{Vikram~Ravi$^1$
\affil{$^1$Cahill Center for Astronomy and Astrophysics MC 249-17, California Institute of Technology, Pasadena CA 91125, USA.; \email{vikram@caltech.edu}}
}

\paperauthor{Vikram~Ravi}{vikram@caltech.edu}{0000-0002-7252-5485}{California Institute of Technology}{Cahill Center for Astronomy and Astrophysics}{Pasadena}{CA}{91125}{USA}

\section{Introduction}

Centimeter-wavelength observations of extragalactic cataclysms and outbursts have resulted in some of the most exciting discoveries in time-domain astronomy. Most recently, the VLA-discovered radio afterglow of the first binary neutron-star merger event to be detected in gravitational waves \citep[GWs;][]{aaa+17} has proven decisive in determining the geometry, energetics and composition of the pre- and post-event ejecta \citep{abf+17,hcm+17,mnh+18}. The VLA also provided the first interferometric localization of a Fast Radio Burst (FRB),  conclusively establishing the distant-extragalactic nature of the phenomenon and identifying an FRB host galaxy \citep{clw+17}. Observations of evolving radio emission associated with the tidal-disruption and accretion of stars by supermassive black holes (tidal disruption events; TDEs) have revealed a diversity of outcomes, from sub-relativistic outflows \citep{abg+16} to relativistic jets \citep[e.g.,][]{zbs+11,mpe+18}. Radio monitoring of the afterglows of $\gamma$-ray bursts \citep[GRBs;][]{fks+01} have enabled accurate calorimetry of the explosions, and the characterization of beaming in GRBs. In turn, the presence or absence of rapidly evolving radio emission has proven to be a crucial discriminant between stellar explosions that result in transient relativistic jets, and those that do not \citep[e.g.,][]{skn+06}. 

This paper outlines the means by which the ngVLA can discover the electromagnetic (EM) counterparts to binary supermassive black holes (SMBHs) caught in the act of coalescence, using their GW emission. The Laser Interferometer Space Antenna \citep[LISA;][]{aab+17}, and its potential companion missions \citep{lcd+16,hw17}, will detect GWs from a few to a few hundred coalescing binary SMBHs each year. Following the success of the LISA Pathfinder mission \citep{aaa+16}, the European Space Agency selected a three-satellite LISA mission for its L3 launch in 2034. From its beginning, LISA will be a bona fide observatory for GW astrophysics. For example, several known ultracompact Galactic white-dwarf binaries will appear like radio-frequency birdies in the $10^{-4}-0.1$\,Hz LISA band \citep{sv06}, and the unresolved background of such binaries will form the dominant broadband LISA ``noise'' source \citep{fp03}. However, LISA detections of inspiralling and coalescing binary SMBHs will probe the wholly unknown formation and growth mechanisms of SMBH seeds, and help unravel the rich astrophysics that governs the fates of binary SMBHs in merging galaxies \citep[for reviews, see][]{h13,c14}. 

Much of the uncertainty in population synthesis models for the binary SMBHs to be detected by LISA reflects our lack of knowledge of the formation and evolution of these systems \citep{kbs+16}. For example, the two leading models for SMBH seeding (remnants of $O(10^{2}M_{\odot})$ Pop. III stars, or $O(10^{5}M_{\odot})$ black holes formed through the gravitational collapse of primordial-gas disks) result in detection rates that vary by an order of magnitude for some LISA configurations. The characteristics of the $10^{-4}-0.1$\,Hz LISA sensitivity curve, combined with the nature of GW emission from binary systems, makes LISA most sensitive to the coalescences of binaries with masses $M_{B}\sim10^{5}-10^{8}M_{\odot}$ at redshifts up to $z\sim10$. GWs from binary SMBHs will sweep through the LISA frequency band for days to years, and localization error regions of $\sim1-10$\,deg$^{2}$ are expected in the days to weeks prior to coalescence \citep{lh08}. 

EM identifications of LISA-detected binaries are required to realize their scientific promise. GW-only detections of SMBH-SMBH coalescence events will supply component masses and redshifts with $O(10\%)$ accuracies \citep{h02}. However, redshifts based on host-galaxy identifications will fully specify the parameters of the coalescing systems, and enable their use as alternative probes of cosmological expansion \citep{tcb+16}. The evolution of the gravitational waveforms of binary SMBHs exactly encode the (redshifted) mass of the system, which in turn provides GW luminosity estimates with no scatter. Thus, redshift measurements of LISA-detected binaries will lead to the assembly of a Hubble diagram with unprecedented accuracy up to $z\sim10$. The characterization of a sample of coalescing-SMBH host galaxies will also provide crucial insight into the environments and mechanisms conducive to the formation and orbital decay of binary SMBHs \citep{c14}. SMBH formation scenarios will be refined \citep{kbs+16}, in a complementary manner to other instruments such as JWST \citep{npf+17}. The nature of the EM signature itself will further test models for interactions between the SMBHs and their environments, such as the formation and sustenance of accretion disks and relativistic jets \citep{s11}. 

In this article, I focus on prompt EM signatures of coalescing binary SMBHs, which will enable contemporaneous multi-messenger studies of LISA detections. I do not consider the possibility of detecting EM signatures of binary-SMBH coalescences {\em independently} of GW observations, because the low event rate and faintness of the potential signatures would necessitate an impracticably large survey.\footnote{Elsewhere in this volume, Burke-Spolaor et al. describe the EM signatures of binary SMBHs prior to coalescence.} In \S2, I describe predictions for relativistic-jet launching upon SMBH-SMBH coalescence, and the estimated radio counterparts. In \S3, I assess the rate of background/interloper events, and conclude in \S4 with expectations for ngVLA observations.

\section{Relativistic jets launched upon SMBH-SMBH coalescence}

Binary SMBHs will likely form in environments rich in dynamically cold gas \citep{kbh17}, and thus be embedded in accretion disks. Prior to coalescence, a binary SMBH will have caused its accretion disk to retreat to a radius where viscous torques in the disk balance the gravitational torques. Periodic accretion episodes will nonetheless result from gas crossing into the hollow center of the disk \citep{rds+11}, and the final violent accretion of this residual gas may cause a flare at $\sim1$ day before the merger \citep{csm+10}. The near-field perturbations of space caused by the coalescing SMBH will partially dissipate into any gas present, resulting in a prompt (tens of minutes) thermal EM flare that can be comparable to the Eddington luminosity of the system \citep{k10}. Rapidly time-variable emission may also be caused by the gravitational recoil of the post-coalescence SMBH perturbing the accretion disk \citep{alm+10}. 

Several groups have simulated the evolution of fossil accretion disks around coalescing SMBHs. \citet{pll10} adopted the force-free approximation for a magnetically dominated tenuous plasma binary-SMBH environment, and found that collimated Poynting-flux outflows were launched from each SMBH during the inspiral phase, with a luminosity of  $10^{43}[M/(10^{8}M_{\odot})]^{2}(v/c)^{2}$\,erg\,s$^{-1}$ that peaked at coalescence. An updated assessment of the Poynting flux emanent from such a system by \citet{mar+12} showed that the dual jets were sub-dominant by a factor of $\sim100$ to a quadrupolar Poynting outflow. However, substantively different results were obtained using full GRMHD simulations of initially matter-dominated accretion flows (plasma $\beta=p_{\rm gas}/p_{\rm mag}=40$) onto coalescing SMBHs \citep{gbm+12,kbe+17}, which likely better represent reality \citep[e.g.,][]{nmn+12}. The capability of the latter simulations to trace magnetic flux-freezing, and resolve MHD turbulence, enabled the compression and amplification of the magnetic field in plasma accreting onto the SMBHs to be identified. Unlike simulations in the force-free approximation, a single collimated Poynting outflow along the angular-momentum vector of the binary was observed, with a few-hour peak in luminosity at the time of coalescence, prior to settling into a higher-luminosity single-SMBH outflow at $\tau_{\rm final}\sim10^{6}[M/(10^{8}M_{\odot})]$\,s after coalescence ($M$ is the initial SMBH mass, assuming an equal mass-ratio system). The magnetic field amplification resulted in a stronger prompt luminosity peak of $L_{\rm prompt}=5\times10^{47}[M/(10^{8}M_{\odot})]^{2}$\,erg\,s$^{-1}$, and a final outflow luminosity of $L_{\rm final}=10^{48}[M/(10^{8}M_{\odot})]^{2}$\,erg\,s$^{-1}$. These quantities are robust to variations in the initial magnetic-field energy density within a reasonable range, but scale with the initial matter density, $\rho_{0}$, as $\rho_{0}/(10^{-11}\,{\rm g\,cm}^{-3})$.  

\begin{figure}[h]
    \centering
    \includegraphics[scale=1.0,trim=0cm 3cm 0cm 1cm,clip]{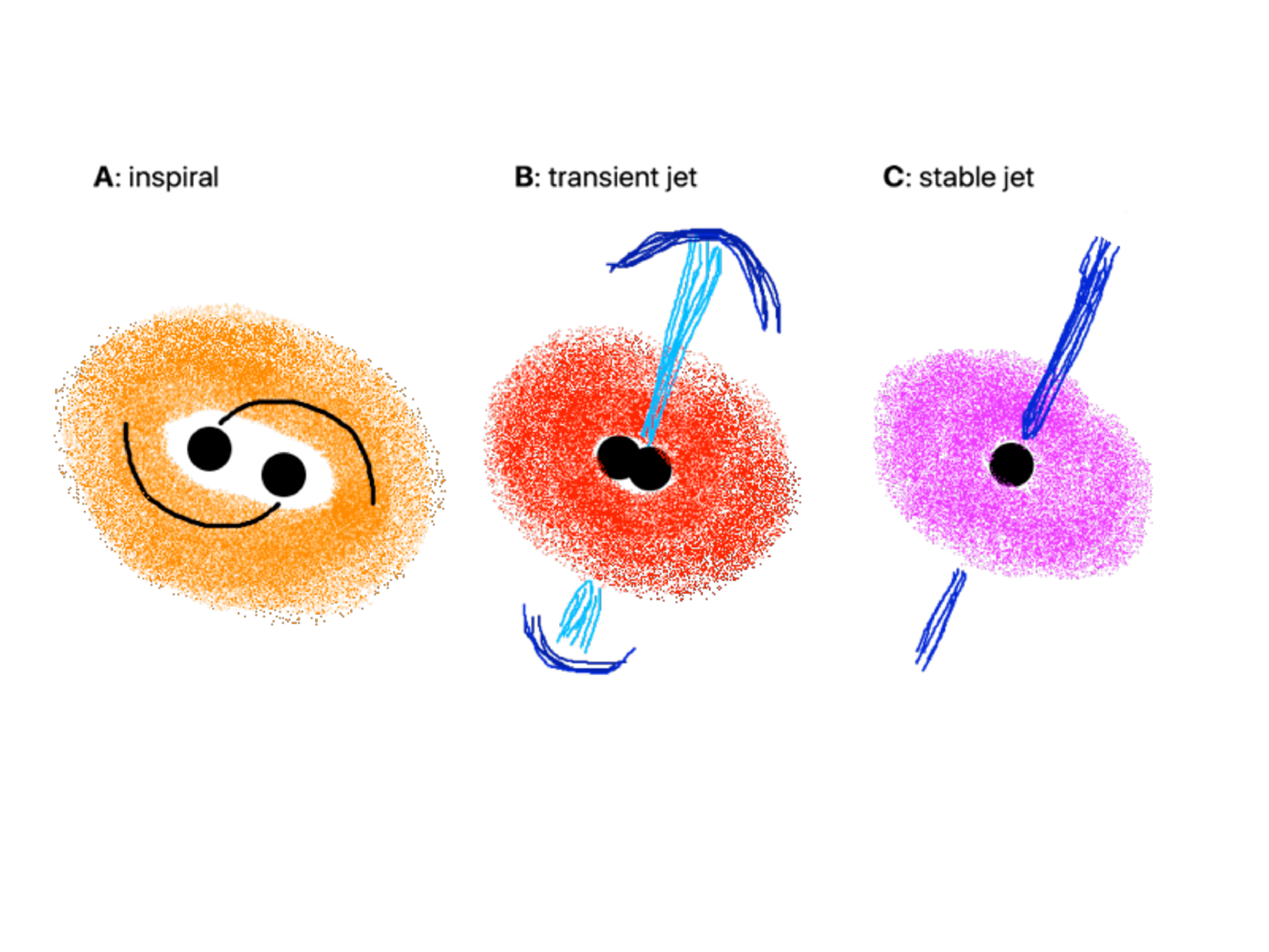}
    \caption{Cartoon illustration of the proposed EM signature of binary-SMBH coalescence events, to be targeted by the ngVLA. {\em Panel A:} LISA will detect and localize inspiralling binary SMBHs in the days to weeks prior to coalescence. {\em Panel B:} Upon coalescence, a prompt transient jet is launched, which drives a shock in the circum-nuclear medium. Radio synchrotron emission from this shock is detectable with the ngVLA. {\em Panel C:} The system eventually settles into a stable state, with a persistent jet that is also possibly radio-detectable.}
    \label{fig:3}
\end{figure}

Collimated Poynting-flux outflows are the leading model for the launching of relativistic astrophysical jets \citep[e.g.,][]{sp11}. I interpret the results of \citet{gbm+12} and \citet{kbe+17} as indicative of the genesis of  relativistic jets upon SMBH-SMBH coalescences in plasma-rich environments. The main uncertainty in the total jet power is the typical matter density in the immediate surrounds of binaries; I assume the presence of a radiatively inefficient accretion flow onto the binary during its final inspiral, with the fiducial value of $\rho_{0} = 10^{-11}$\,g\,cm$^{-3}$. The mechanisms by which initially Poynting-dominated jets attain their mass loading, and hence their finite Lorentz factors $\Gamma_{0}=(1-\beta^{2})^{-1/2}$, are unknown, as are the characteristic length scales on which this occurs \citep{sbm+05}. However, based on VLBI observations of superluminal AGN jets, I adopt $\Gamma_{0}=10$ \citep{hvt+09}, and assume that particle entrainment occurs on sub-parsec scales \citep{pk15}. Assuming full efficiency in the conversion of Poynting flux to jetted matter/energy, I model the Poynting-luminosity increase coincident with SMBH-SMBH coalescence as a $\tau_{\rm prompt}=10^{3}[M/(10^{8}M_{\odot})]$\,s transient jet with a total energy output of $E_{\rm prompt} = 5\times10^{50}[M/(10^{8}M_{\odot})]^{3}$\,erg. The properties of the final jet ($\tau_{\rm final}$ and $L_{\rm final}$) are as above. The model is schematically illustrated in Fig.~\ref{fig:3}. 

Relativistic jets have several observational manifestations. In this paper, I predict their centimeter-wavelength radio signatures. The persistent twin jets observed in the binary-SMBH simulations of \citet{pll10} led \citet{kos+11} and \citet{oks+11} to model the associated radio emission, assumed to be due to relativistic electrons advected with each jet, simply as a fixed fraction of the total jet luminosity. However, the transient nature of the jet model that I adopt here necessitates a different approach.

\subsection{Predictions for centimeter-wavelength transients}

Radio emission due to relativistic jets and outflows is generated by electrons emitting synchrotron radiation. For on-axis AGN (e.g., blazars, BL Lacs), a correlation exists between the bolometric luminosity $L_{\rm jet}$ (for which the $\gamma$-ray luminosity is a rough proxy) and the radio-synchrotron luminosity $L_{\rm rad}\sim\nu L_{\nu}$, where $\nu$ is the radio frequency and $L_{\nu}$ is the radio spectral luminosity at its approximate peak \citep{ggt+11}. The bolometric luminosity can in turn be estimated using the total jet kinetic power $P_{\rm jet}$ \citep{ngg+12}. I use these empirical results to estimate the peak radio flux density of the persistent final jet associated with a binary-SMBH coalescence event as
\begin{equation}
F_{\rm final} \approx 15a^{3}[M/(10^{8}M_{\odot})]^{2}[D_{L}/(45\,{\rm Gpc})]^{-2}\,{\rm mJy},
\label{eqn:1}
\end{equation}
where $a=(1-\beta)/(1-\beta\cos\theta)$ ($\beta=(1-\Gamma^{-2})^{-1/2}$) is a factor that accounts for an off-axis observer orientation at an angle $\theta$,  $D_{L}$ is the luminosity distance, and I assume $\nu_{p}=10$\,GHz for the rest-frame spectral peak.\footnote{Compact radio sources with spectral peaks in the tens of GHz range are thought to be the youngest instances of active galaxies \citep{o98}. This motivates our fiducial choice of $\nu_{p}=10$\,GHz for the spectral peak of the persistent final jet.} Equation~\ref{eqn:1}, and the assumption of a constant-$\Gamma$ jet, implies that the final jet is only important for small viewing angles $\theta$. For example, $\theta=60^{\circ}$ implies $a=0.01$, reducing the fiducial $F_{\rm final}$ to $15$\,nJy. Note that the GW strain amplitudes emitted by face-on binaries are a factor of four larger than those emitted by edge-on binaries, and that small values of $\theta$ are therefore more likely to be observed \citep{w87}. 

The $L_{\rm jet}-L_{\rm rad}$ correlation, which is stronger when only variable emission is considered (albeit with a time-lag), is interpreted as all the EM emission originating from particles accelerated in internal shocks within jets. In the case of newly launched transient jets such as in $\gamma$-ray bursts, particle acceleration also occurs in the external shock at the interface between the jet and the circum-nuclear medium \citep[CNM;][]{p04}. The initial transient jet will drive a relativistic shock through the CNM, which becomes Newtonian once the kinetic energy of the swept-up CNM is equivalent to that of the jet. The final jet will in turn power a shock within the initial jet, with the radio luminosity  estimated above, that will ultimately further accelerate the forward CNM shock.

\begin{figure}
    \centering
    \includegraphics[scale=0.7,angle=-90]{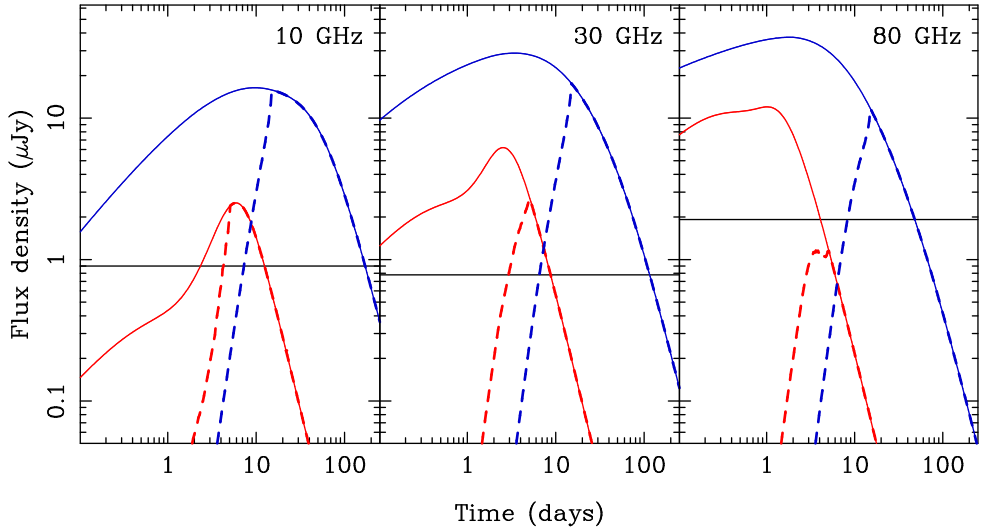}
    \caption{Predicted lightcurves at 10\,GHz, 30\,GHz and 80\,GHz of emission corresponding to the external CSM shock due to the prompt transient jet predicted to occur during SMBH-SMBH coalescence events. The blue (higher-flux) curves correspond to the merger of two $10^{8}M_{\odot}$ SMBHs at $z=5$, and the red (lower-flux) curves correspond to the merger of two $10^{7}M_{\odot}$ SMBHs at $z=1$. The solid curves were calculated for on-axis observers, while the dashed curves were calculated for a viewing angle of $60^{\odot}$ to the jet direction. The horizontal lines are the $6\sigma$ detection thresholds for 9\,h ngVLA observations in each band (ngVLA Memo \#5).}
    \label{fig:1}
\end{figure}

I model the radio emission associated with the external CNM shock using the semi-analytic calculations of \citet{lvm+12} implemented in their {\em Spherefit} code. Although these results were derived for a spherical outflow, they are applicable to an on-axis observer of a relativistic jet. To evaluate the off-axis emission, I apply straightforward relativistic corrections relevant to a point-mass in linear motion. The shock Lorentz factor scales approximately as $\Gamma_{0}(t/\tau_{\rm prompt})^{-(3-k)/2}$ in the initial relativistic phase, where $k$ is the index of a power-law describing the radial variation in the CNM density: $\rho(r)\propto r^{-k}$. For an off-axis observer, the time is given by $t/a$, the observed frequency is given by $\nu a$, and the observed flux-density is given by $a^{3}F_{\nu}$. I neglect any contributions to the observed radio emission from a reverse shock propagating backwards through the transient jet. Observations of stellar tidal disruption events \citep[e.g.,][]{abg+16} motivate a fiducial CNM number-density profile of $\rho(r) = 10[r/(10^{17}\,{\rm cm})]^{-1.5}$\,cm$^{-3}$. Finally, I make the standard assumption for the accelerated electron power-law energy spectral index of $p=3$. Predicted lightcurves for the external-shock emission are shown in Fig.~\ref{fig:1}. 

Emission in EM bands besides the radio is unlikely to be of significant importance for the jet model considered here, unless the jet is viewed on-axis. Synchrotron emission from the CNM shock will be most readily detected in the radio band \citep{spn98}. Relativistic jets powered by SMBHs are also most easily detected in the radio band at high redshifts \citep[e.g.,][]{md08}. The wider landscape of (non-jetted) EM signatures of SMBH-SMBH coalescence is a topic of intense investigation \citep{s11}. Nonetheless, in analogy with TDEs such as that observed in the galaxy merger Arp\,229 \citep{mpe+18}, dust obscuration may significantly affect signatures in important wavelength ranges (optical, soft X-ray) besides the radio. 

\section{The rate of background events}

\begin{figure}[h]
    \centering
    \includegraphics[scale=0.4,angle=-90]{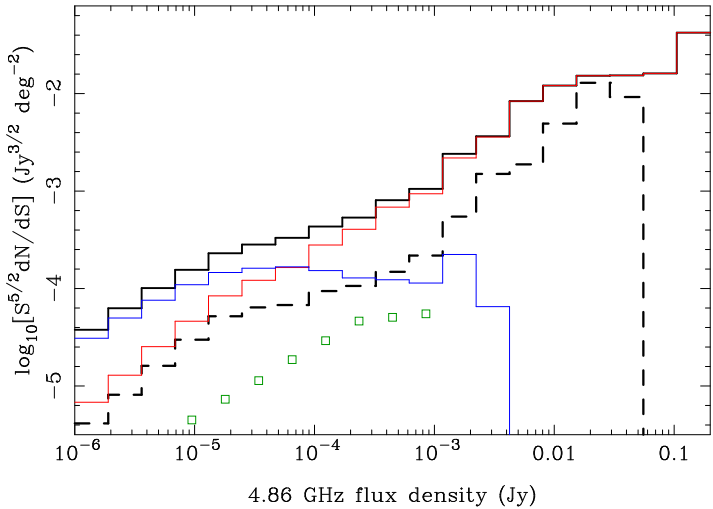}
    \caption{Simulated radio-source differential counts at 4.86\,GHz from the SKA Simulated Skies project \citep{wmj+08} (see http://s-cubed.physics.ox.ac.uk). The thick black trace corresponds to all sources, and the thick dashed black trace indicatessources larger than 50\,mas. The ngVLA will thus be able to refine transient/variable candidate lists by factors of $\gtrsim4$ in a single epoch by excluding extended sources. The green squares correspond to a simulation of the maximum possible counts of sources with angular extents $<5\,\mu$as, which are likely to refractively scintillate at moderate to high Galactic latitudes (Ravi et al., in prep). Note that the most common cataclysmic-event interloper, core-collapse supernovae, will have $S^{5/2}dN/dS\approx2\times10^{-7}$\,Jy$^{3/2}$\,deg$^{-2}$ \citep{mhb+16}.}
    \label{fig:2}
\end{figure}

The ngVLA will need to survey GW localization regions of several deg$^{2}$ to detect radio counterparts to binary-SMBH coalescences. Radio emission associated with prompt transient jets will vary on timescales of days to weeks, while emission associated with persistent final jets will turn on over weeks to months. It is important therefore to understand the background rate of transient and variable radio sources with flux densities and variability timescales comparable to the objects of interest. Variability in the radio sky is not empirically characterized at these faint flux densities. Below, I consider in turn the contributions from cataclysmic events, scintillating compact AGN, and intrinsic variability in AGN. 

\textbf{Cataclysmic events.} The afterglows of core-collapse supernovae are the most common cataclysmic events in the radio sky \citep[for a compilation, see][]{mhb+16}. Assuming flux-density statistics consistent with a non-evolving population in Euclidean space, 6\,deg$^{-2}$ events are expected peaking over 10\,$\mu$Jy. However, besides the relativistic explosions that form $\sim1\%$ of the supernova population, the variability timescales of radio supernovae will be one to two orders of magnitude larger than the prompt binary-SMBH coalescence counterparts. Furthermore, the low radio luminosities of radio-supernova afterglows imply a nearby population even for the ngVLA, at redshifts $z\lesssim0.3$. Approximately $0.01-0.1$\,deg$^{-2}$ on-axis GRBs and jetted tidal disruption events are expected, which will more closely mimic the predicted prompt counterparts. Thus, at most $\sim1$ relativistic explosion is expected to form an interloper in ngVLA observations with 10\,$\mu$Jy sensitivity of any given 10\,deg$^{2}$ binary-SMBH coalescence localization region. The host galaxies and redshifts of each interloper radio event will need to be identified, and compared with coarse redshift information from the GW detection, to isolate the true binary-SMBH counterpart. Accurate radio localization to $\sim10$\,mas (to match Gaia-based optical astrometric accuracy) may also enable off-nuclear sources to be discarded.

\textbf{Scintillating AGN.} Spatio-temporal density variations in the Milky Way ionized-ISM cause refractive scintillations of compact extragalactic radio sources. Maximum modulations (modulation indices of order unity) are observed for sources smaller than a few $\mu$as, at frequencies between $1-10$\,GHz off the Galactic plane \citep{w98}, on timescales of several hours. Although compact sources (e.g., observable with VLBI) are a subdominant population below flux densities of 1\,mJy \citep{mdn+13}, any optically thick synchrotron source fainter than 1\,mJy at a few GHz will compact enough to scintillate. Simulations (Ravi et al., in prep) suggest that up to 500\,deg$^{-2}$ scintillating sources $>10$\,$\mu$Jy will be observable at $10$\,GHz. These sources can be distinguished from the prompt counterparts of binary-SMBH coalescences by (a) higher-frequency observations, where modulation indices will typically be lower, (b) wide-band monitoring to identify unusual instantaneous spectral shapes caused by scintillation \citep[e.g.,][]{bst+16,alb+17}, and (c) longer-term monitoring to identify re-brightening episodes. 

\textbf{Variable AGN.} Intrinsic variability in compact AGN may be a more insidious interloper class. \citet{mhb+16} suggest that $\sim1$\,deg$^{-2}$ AGN is expected to vary on timescales of days to years above $0.3$\,mJy, possibly due to the propagation of shocks internal to jets \citep{mg85}. This implies the existence of a large interloper population at the tens of $\mu$Jy level, given the still-increasing AGN source counts in this flux-density range \citep[e.g.,][and Fig.~2 here]{ccf+12}. Having discarded all interlopers that are not temporally coincident with the GW events, and those that are outside the GW-derived redshift bounds, the only way to distinguish between an internal shock within a pre-existing jet and a newly formed jet shocking an external CNM is by careful modeling of the evolution of the radio source. Such modeling could, for example, distinguish between a shock propagating through a dense, radially varying CNM, and a sparse jet.

\section{Localizing LISA events with the ngVLA}

Radio observations will continue to be of great value in classifying and characterizing enigmatic classes of extragalactic transients in the ngVLA era. Dedicated ngVLA observations of the few to a few hundred coalescing binary SMBHs to be detected annually by LISA are required for the discovery of their EM counterparts. The localization of the EM counterparts will result in their host galaxies and redshifts being identified, and will unlock their rich astrophysical and cosmological potential. LISA will provide localization regions of $1-10$\,deg$^{2}$ in the days to weeks prior to coalescence, and redshift measurements accurate to $O(10\%)$. GRMHD simulations of inspiralling SMBHs embedded in realistic accretion flows predict a prompt jet with energy $E_{\rm prompt} = 5\times10^{50}[M/(10^{8}M_{\odot})]^{3}$\,erg lasting a few hours upon coalescence. The systems will then, on timescales $\tau_{\rm final}\sim10^{6}[M/(10^{8}M_{\odot})]$\,s, launch stable jets with lumonisities $L_{\rm final}=10^{48}[M/(10^{8}M_{\odot})]^{2}$\,erg\,s$^{-1}$. Examples of the radio lightcurves expected due to the prompt transient jets shocking the CNM are shown in Fig.~\ref{fig:1}, and the final jet radio flux density is estimated in Equation~\ref{eqn:1}. For example, for a fiducial off-axis observer orientation, coalescences of $10^{8}M_{\odot}$ SMBHs at $z=5$ will produce prompt jets with radio flux densities peaking above $10\,\mu$Jy at 10, 30, and 80\,GHz, within $\sim10$ days of coalescence. The final jets will generally be observable only in more favorable orientations, depending on their Lorentz factors. 

From an observational point of view, I have provided specific motivation for the use of the ngVLA to search for individual faint transient/variable sources in up to few-deg$^{2}$ regions. Some general technical considerations for this task are discussed below.

\begin{description}

\item[Point-source sensitivity.]  The mooted performance of the ngVLA at centimeter wavelengths, in particular above the 14\,GHz limit of the baseline design for SKA1-mid, is highly desirable. Although lower frequencies (1--10\,GHz) are typically used for time-domain work, synchrotron emission from expanding CNM shocks will have self-absorption spectral peaks that are brighter at higher frequencies at earlier times. Centimeter-wavelength observations therefore enable the proposed LISA-event counterparts to be detected sooner, and allow for the evolution of the self-absorbed spectrum to be monitored, thus characterizing the expansion velocity, total energy, and the CNM density profile. 

\item[Survey speed.] The ngVLA as defined in Memo \#5 will require several pointings to survey few-deg$^{2}$ regions for transient/variable sources. For the LISA case, observations to depths of a few $\mu$Jy will be required. The approximate, optimistic ngVLA survey speeds corresponding to $5\,\mu$Jy rms continuum noise in the 2, 10, 30, and 80\,GHz bands are 5.3, 0.9, 0.14, and 0.002 deg$^{2}$/hr. Dwell times of $20-130$\,s per pointing are required, which may motivate an on-the-fly mosaicking approach in some cases. However, it is evident that deep cadenced surveys of few-deg$^{2}$ regions will only be feasible with the ngVLA in the 2\,GHz and 10\,GHz bands, and possibly the 30\,GHz band in exceptional cases. In cases where rms noise levels of $\gtrsim10\,\mu$Jy are sufficient, the dwell times per pointing become small enough that sub-arraying to cover a larger frequency range becomes a possiblity. Otherwise, higher-frequency follow-up of individual sources of interest detected at lower frequencies will be a more practicable strategy. 

\item[Angular resolution and astrometric accuracy.] Wide-field imaging with the highest angular resolution and corresponding astrometric accuracy achievable with the ngVLA is desirable for this science case. The majority of sources near the detection threshold of a naturally weighted image with sub-$100$\,mas angular resolution and few-$\mu$Jy sensitivity will be star-forming galaxies and extended \citep[e.g.,][]{ccf+12}, and thus separable from the point-like sources of interest. This is demonstrated in Fig.~\ref{fig:2}, where I compare simulated source counts of $<50$\,mas objects with the total radio-source population.  Astrometric accuracy of $\lesssim10$\,mas will further allow for accurate radio-optical image registration in the post-Gaia era, enabling the rejection of some interloper radio-transient events. 

\item[Spectral coverage.] The characterization of any detected LISA-event counterpart over the full ngVLA band would be an ideal outcome. In particular, it is important to identify and monitor any continuum spectral peak. Wide-band  data are also useful in rejecting interloper events. For example, scintillation of compact sources is mitigated at higher frequencies (typically $\gtrsim10$\,GHz), and can sometimes be identified by unusual spectral shapes. In-band spectral indices will also be useful in source identification. 

\item[Triggered, cadenced observing.] Finally, the ngVLA use-case presented here will require support for triggered, cadenced observing. Procedures for such observing modes have been honed on the VLA. However, triggered observations have been more difficult to implement with ALMA \citep{abb+17}. It is important that policies enabling fast-turnaround proposals and rapidly scheduled observations are included in ngVLA operations planning. Further, tools that enable rapid data reduction  by the user community will also be necessary. 

\end{description}

\acknowledgements I thank S. Phinney, S. Kulkarni, G. Hallinan, J. Lazio, and S. Burke-Spolaor for useful discussions, and acknowledge funding from the ngVLA Community Studies Program.   


\end{document}